Detection of Individual Vortices in Micron-Size $Sr_2RuO_4$ Rings by Phase-Locked Cantilever Magnetometry

Joonho Jang, Raffi Budakian[a]

Department of Physics, University of Illinois at Urbana-Champaign, Urbana, IL 61801-3080

Yoshiteru Maeno

Department of Physics, Kyoto University, Kyoto 606-8502, Japan

**Abstract**

We describe a feedback-based dynamic cantilever magnetometry technique capable of achieving high magnetic moment sensitivity with low applied fields. Using this technique, we have observed periodic entry of vortices into mesoscopic $Sr_2RuO_4$ rings. The quantized jump in the magnetic moment of the particle produced by individual vortices was measured with a resolution of $7 \times 10^{-16}$ *e.m.u.* at an applied field of 1 Oe.

In the past decade, there has emerged strong evidence to support p-wave symmetry in the layered-perovskite superconductor $Sr_2RuO_4$, whose ground state is thought to be analogous to the A-phase of $^3$He [1-4]. It is believed that the spin and orbital degrees of freedom of the superconducting order parameter can give rise to states with remarkable properties, such as chrial domains [5] and half-quantum vortices (HQVs) that may obey non-Abelian statistics [6]. These states can be identified by their magnetic response. For a $p_x \pm ip_y$ pairing state that breaks time reversal symmetry, domains having a particular chirality should possess edge currents and a corresponding magnetic moment. In the case of HQVs, the magnetic moment produced by the orbital phase winding should be half that of conventional full quantum vortices. With regards to the latter, recent theoretical work suggests that HQVs could be made energetically favorable in mesoscopic annular geometries whose dimensions are comparable to the penetration depth [7].

---

[a] Email: budakian@illinois.edu

Detecting the weak magnetic response from micron-size superconductors requires highly sensitive probes. SQUID and Hall magnetometry are ubiquitous in sensitive magnetic measurements and have been successfully used to study superconductivity in mesoscopic aluminum disks [8] and rings [9]. Cantilever magnetometry has also been used to achieve outstanding sensitivity in detection of magnetic fluctuations in nanometer size ferromagnetic particles [10], observation of mesoscopic vortex physics in superconductors [11], and has recently been proposed to study persistent currents in mesoscopic normal metal rings [12].

Cantilever-based detection offers higher sensitivity and a cleaner electromagnetic environment than other approaches. In most applications, however high detection sensitivity is achieved by applying large magnetic fields. Such an approach is not well suited for determining the low-field magnetic response of the sample. In this paper, we introduce a new method of cantilever magnetometry which achieves high detection sensitivity with low applied fields. Using this technique, referred to as phase-locked cantilever magnetometry, we have observed individual vortices in micron-size $Sr_2RuO_4$ rings with a resolution of $7 \times 10^{-16}$ *e.m.u.* with only 1 Oe amplitude modulated field applied to the sample.

Samples are fabricated from high quality Sr$_2$RuO$_4$ crystals with $T_c = 1.43\,K$ grown using the floating-zone method [13]. Micron-size pieces of Sr$_2$RuO$_4$ are obtained by crushing millimeter size single crystals and imaging the pieces using the electron beam of a dual-column focused ion beam (FIB.) Sr$_2$RuO$_4$ is highly micaceous and easily cleaves parallel to the *ab* planes. To fabricate the annular geometry, a focused beam of gallium ions is used to cut a hole through the center of the particle in the direction parallel to the c-axis. The particle is then glued to the tip of a custom-fabricated silicon cantilever with the crystal c-axis oriented perpendicular to the cantilever. Figure 1 shows the experimental setup. Because of the odd spatial parity of Sr$_2$RuO$_4$, very small amounts of non-magnetic impurities are known to destroy superconductivity [14]. We found, by minimizing the exposure to gallium ions during the FIB process, the critical temperature of the ruthenate particles could be maintained within 0.3 K of the bulk value.

In the presence of an external field $\vec{H}$, the magnetic response of an annular superconducting particle has two contributions: (1) Meissner currents that expel flux from the interior of the superconductor and (2) currents that ensure the fluxoid enclosed by the hole is quantized. In the low field limit, the total magnetic moment will be the sum of these two contributions $\vec{\mu} = \vec{\mu}_M + \vec{\mu}_V$, where $\vec{\mu}_M = \chi \vec{H}$ and $\vec{\mu}_V = n(\vec{H})\Delta\vec{\mu}_V$ [15]. Here, $\chi$ is the diamagnetic susceptibility tensor, $\Delta\vec{\mu}_V$ is the change in moment associated with vortex entry and $n(\vec{H})$ is the winding number of the order parameter whose equilibrium value depends on $\vec{H}$. In this paper, we measure the out-of-plane component of the magnetic moment $\mu_z$ associated with fluxoid quantization using dynamic mode cantilever magnetometry inside a $^3$He refrigerator with a base temperature of 300 mK.

In dynamic detection, one measures the change in the cantilever frequency or dissipation in response to the externally applied field. In our setup, the displacement of the cantilever is measured using a fiber optic interferometer operating at 1510 nm. The optical power is maintained at 5 nW to minimize heating of the sample caused by optical absorption. During measurement, the cantilever is placed in a positive feedback loop and is driven at its natural frequency $\omega_c$ using a piezoelectric transducer; the resulting tip motion is given by $z(t) = z_{pk} \cos \omega_c t$. A digital controller implemented in a field programmable gate array (FPGA) maintains a constant tip amplitude $z_{pk}$ and is also used to determine both the instantaneous cantilever frequency and dissipation.

The flux through the ring is varied by applying a static field $H_z$. In addition, the magnetic field $H_x(t) = \Delta H_x \cos(\omega_c t + \theta_x)$ is applied parallel to the $x$-direction that is phase-locked to the cantilever position $z(t)$ (see Fig. 1); $\Delta H_x$ is the modulation amplitude and $\theta_x$ is used to adjust the relative phase between the applied modulation and the cantilever motion. $H_x(t)$, which depends parametrically on $z(t)$, is derived from the phase-locked loop (PLL) controller implemented in the FPGA that tracks the instantaneous phase of the oscillator. In the presence of external fields, the magnetic moment $\vec{\mu}$ produced by superconductor generates a torque $\vec{\tau} = \vec{\mu} \times \vec{H}$ which acts on the cantilever. By expanding the $z$-component of the resulting force, and keeping terms linear in the cantilever displacement, the Fourier transform of the linearized equation of motion for the cantilever is obtained.

$$(-\omega^2 + i\gamma\omega + \omega_0^2)\tilde{z}(\omega) \approx \frac{\omega_0^2}{kL_e^2} \mu_z \left\{ H_z + \Delta H_x e^{i\theta_x} \left( \frac{L_e}{z_{pk}} \right) \right\} \tilde{z}(\omega) \qquad (1)$$

Here, $\omega_0$, $k$ and $\gamma$ are the cantilever frequency, spring constant and dissipation in the absence of magnetic interactions; $L_e \approx L/1.38$ is the effective length of the cantilever. The first term on the right-hand-side of Eq.(1), which we refer to as the passive dynamic force, relies on the bending of the cantilever to generate an oscillating magnetic field $H_x \approx H_z z / L_e$ in the reference frame of the superconductor; the resulting position-dependent force shifts the cantilever frequency [10]. Expressing the field dependence of $\mu_z$ indicated above and assuming the c-axis susceptibility $\chi_z = \partial \mu_z / \partial H_z$ to be constant gives $\Delta f_P \approx f_0 (\chi_z H_z^2 + \Delta \mu_z n H_z) / 2kL_e^2$, where $f_0 = \omega_0 / 2\pi$ and $\Delta \mu_z$ is the $z$ component of the jump in moment caused by vortex entry. Note, in writing Eq. (1) we have assumed that $\chi_z \gg \chi_x$ and neglected the contribution to the torque from $\chi_x$. This assumption is justified because of the large shape anisotropy of our sample as well as the highly anisotropic response of the superconductor. In Sr$_2$RuO$_4$, $\lambda_{ab}(0) = 152\,nm$, $\lambda_c(0)/\lambda_{ab}(0) \approx 20$, where $\lambda_{ab}(0)$ and $\lambda_c(0)$ refer to the zero temperature in-plane and c-axis penetration depths.

Dynamic magnetometry offers a number of advantages. First, operating at the resonant frequency enhances displacement sensitivity of the oscillator by the quality factor $Q$ and greatly facilitates thermal limited force detection. Second, variations in the cantilever frequency caused by changes in $\mu_z$ can be tracked instantaneously. If we calculate the magnetic moment sensitivity, however we find that the spectral density of moment fluctuations for passive dynamic detection $\left(\sqrt{S_\mu}\right)_{passive} = \left(\sqrt{S_\mu}\right)_{static} (L_e / z_{pk})$ is larger by a factor of $L_e / z_{pk}$ than for static torque measurement $\left(\sqrt{S_\mu}\right)_{static} = \sqrt{S_F} L_e / H_z$; here, $S_F = 4k_B Tk / \omega_0 Q$ is the spectral density of thermal force fluctuations. For our experimental parameters $L_e \approx 55\,\mu m$ and $z_{pk} \approx 60\,nm$, the reduction in sensitivity $z_{pk}/L_e \sim 10^{-3}$ is significant. Phase-locked magnetometry overcomes this limitation by using an active approach to directly apply a position-dependent field. Thus, for the same magnitude of the applied field, the detection sensitivity is enhanced by a factor of $L_e / z_{pk}$ relative to passive dynamic measurement.

Figure 2(a) shows passive dynamic magnetometry data ($\Delta H_x = 0$), taken at $T = 0.6\,K$ for a ruthenate particle having dimensions 5 μm ×4 μm ×2 μm with a 1.8 μm diameter hole in the center. For both parts (a) and (b) of the measurement, the equilibrium state of the superconductor is realized by increasing the laser power to momentarily heat the particle above $T_c$ at each value of $H_z$. We find that the frequency response in Fig. 2(a) exhibits a quadratic dependence on $H_z$ indicative of the Meissner response, however the jumps associated with vortex entry cannot be resolved at low fields.

Figure 2(b) shows the same field sweep taken using phase-locked detection. For this data $\theta_x = 0$, and the cantilever frequency shift is given by $\Delta f_{PL} \approx f_0(\chi_z H_z + \Delta \mu_z n)\Delta H_x / 2kL_e z_{pk}$. The data, taken with 1.3 Oe modulation amplitude, clearly shows the linear Meissner response as well as the quantized jumps in magnetization related to periodic entry of vortices into the hole. To estimate the magnitude of the jump, we consider the change in magnetic moment of the particle $\vec{\mu} = \frac{1}{4\pi}\int(\vec{B}-\vec{H})dV$ upon vortex entry. For this annular particle, whose radius $R$, wall thickness $d$ and length $L$, are all large compared to $\lambda_{ab}$, the change in magnetic moment can be easily estimated $\Delta \mu_z \sim R^2 L \Delta B / 4 = \phi_0 L / 4\pi$, where $\Delta B$ is the change in magnetic field inside the hole and $\phi_0 = hc/2e$ is the flux quantum. Based on the dimensions of the ring, the estimated jump $\Delta \mu_z \sim 3 \times 10^{-12}$ e.m.u. is in reasonable agreement with the measured value $\Delta \mu_z \sim 5 \times 10^{-12}$ e.m.u.. From the slope of the data between jumps, we find $\chi_z = 3.8 \times 10^{-12}\,cm^3$.

We also calculate the average slope over the ±4.5 Oe range of the data, which gives $\bar{\chi}_z = 1.1 \times 10^{-12}\ cm^3$; this value is in good agreement with the value $\bar{\chi}_z = 1.2 \times 10^{-12}\ cm^3$ determined from the quadratic fit to the data in Fig. 2(a).

To be more in the mesoscopic regime, i.e. $R, L$ and $d$ comparable to $\lambda_{ab}$, we fabricated a second smaller ruthenate ring having dimensions 1.8 μm ×1.5 μm ×0.3 μm with a 0.74 μm diameter hole. Figure 3 shows data taken by cyclic field sweep starting from $H_z = 0$ at a constant temperature of $T = 0.5\ K$. Phase-locked measurements were obtained in two modes: (a) $\theta_x = 0$ (frequency shift) and (b) $\theta_x = \pi/2$ (dissipation shift.) For mode (b), the shift in dissipation and the magnetic moment are related by $\Delta\gamma_{PL} \approx -\omega_0(\chi_z H_z + \Delta\mu_z n)\Delta H_x / kL_e z_{pk}$. Notice that the measured magnetization is nearly identical for both modes. Mode (b) has the added benefit in that it allows the passive and phase-locked signals to be separately measured. From the standard deviation of the data in Fig. 3(b), we determine the moment sensitivity within the $\tau_m = 1s$ integration time of the measurement to be $\Delta\mu \sim 7 \times 10^{-16}\ e.m.u.$. Based on the cantilever parameters, we find the measured moment sensitivity is near the theoretical limit

$$\Delta\mu = \sqrt{S_F / 2\tau_m} L_e / \mu_0 \Delta H_x \sim 5 \times 10^{-16}\ e.m.u..$$

In brief, we have developed a new ultrasensitive approach to dynamic cantilever magnetometry that achieves thermal-limited moment sensitivity with low applied fields. Using this phase-locked detection scheme, we have observed individual vortices in micron-size rings of $Sr_2RuO_4$. Future work will focus on elucidating the novel states that could exist in mesoscopic samples of this unconventional superconductor.

We thank Dale Van Harlingen, David G. Ferguson, Victor Vakaryuk and Suk Bum Chung for valuable discussions. This work was supported by U.S. Department of Energy Office of Basic Energy Sciences, grant DEFG02-07ER46453 through the Frederick Seitz Materials Research Laboratory at the University of Illinois at Urbana Champaign.

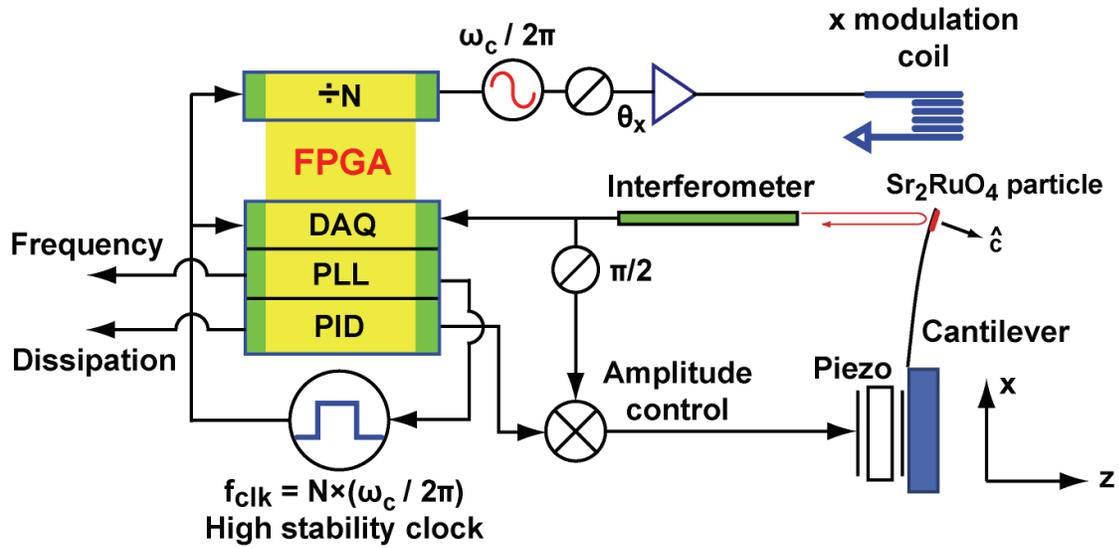

**Figure 1:** Schematic of phase-locked magnetometry setup. The single crystal silicon cantilevers used for measurement are 100-nm thick, 80-μm long and exhibit a thermal-limited force sensitivity of $\sqrt{S_F} \approx 1.0 \times 10^{-18}\ N/\sqrt{Hz}$ at $T = 0.5\ K$.

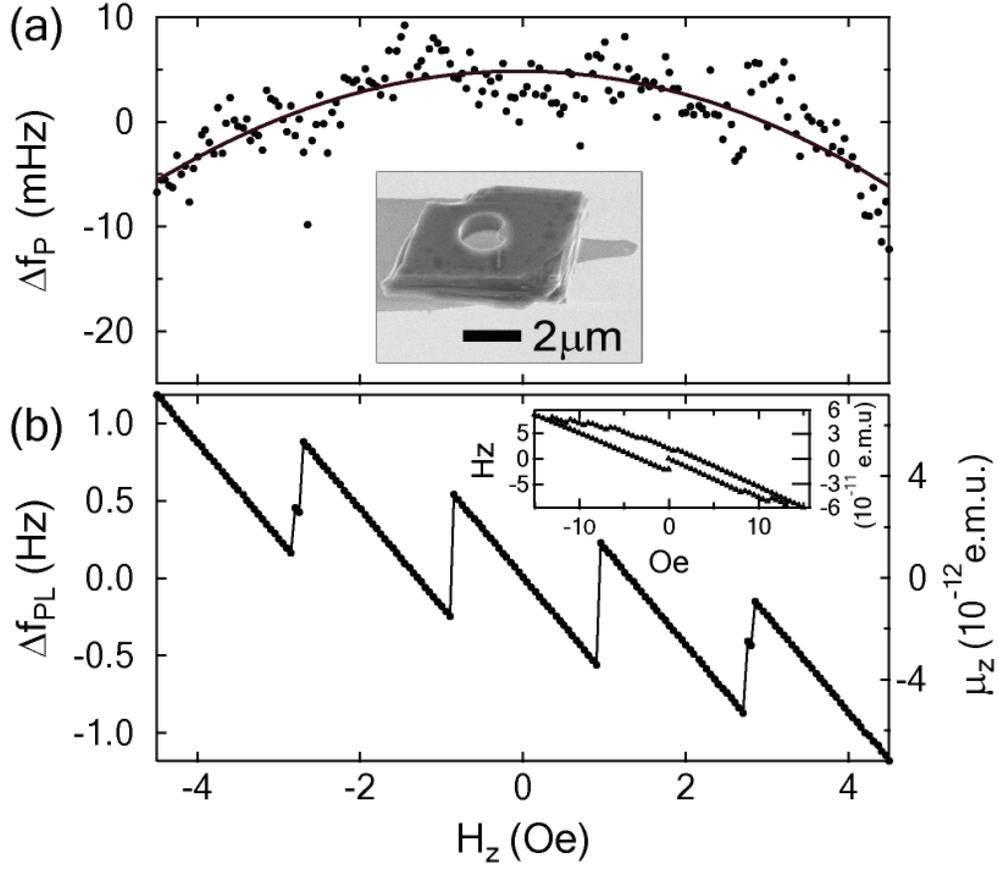

**Figure 2:** Comparison between (a) passive and (b) phase-locked magnetometry data. Each data point for (a) and (b) is a 2 s average. Technically, the data in (b) has a contribution from $\Delta f_P$. Since $\Delta f_{PL} \gg \Delta f_P$ however, we ignore the contribution from $\Delta f_P$ to the frequency shift. The inset in Fig. 2(b) shows data obtained at $T = 0.6\,K$ without temperature cycling the superconductor above $T_c$. The hysteretic behavior is caused by the finite surface barrier for vortex entry which becomes large for rings that have $d > \lambda_{ab}$ [16]. The cantilever parameters for this measurement were $f_0 = 6.1\,kHz$, $k = 2.3 \times 10^{-4}\,N/m$ and $Q = 70,000$.

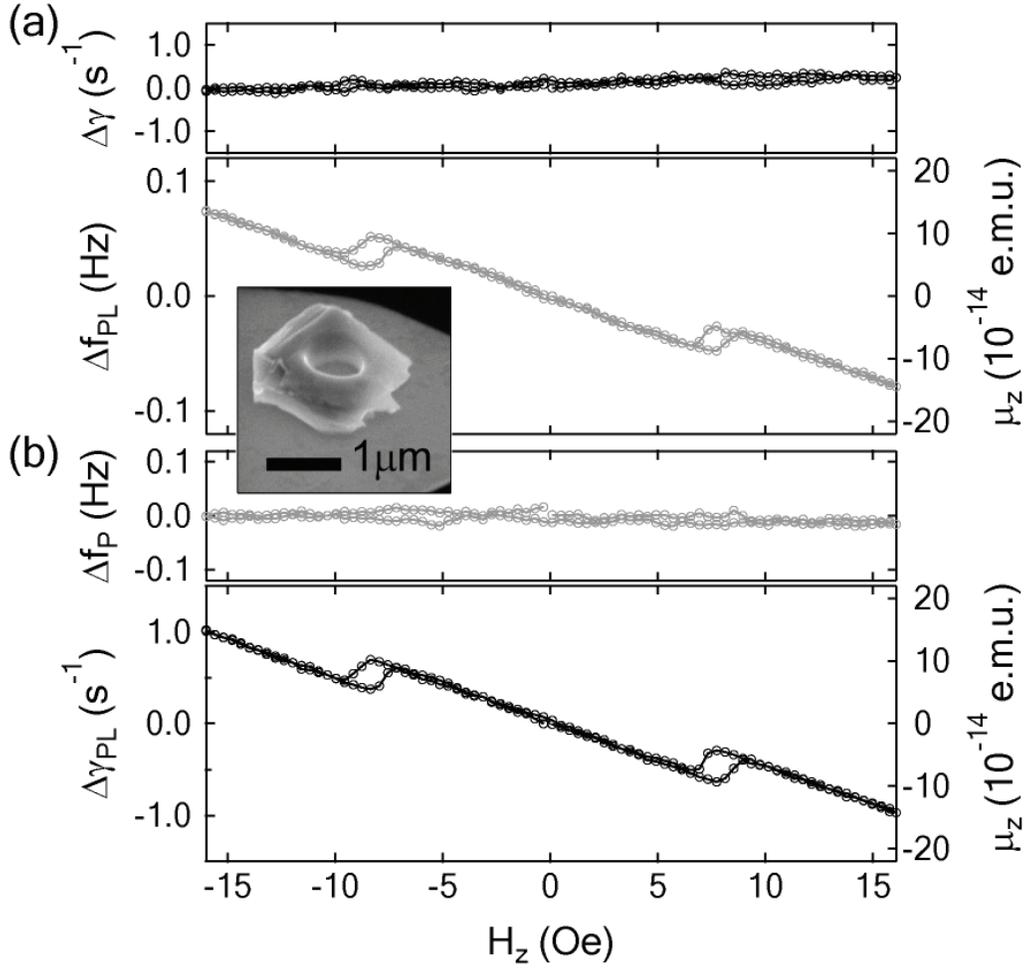

**Figure 3:** Phase-locked detection measurements for the smaller ruthenate particle (inset.) Frequency and dissipation data taken at $T = 0.5\,K$ for (a) $\theta_x = 0$ and (b) $\theta_x = \pi/2$. Once again in (a), we ignore the small contribution from $\Delta f_P$ to the frequency shift. The cantilever parameters for this measurement were $f_0 = 16\,kHz$, $k = 3.6 \times 10^{-4}\,N/m$ and $Q = 65,000$.